\documentclass[a4paper]{VisionStyle}
\usepackage{epsfig}

\advance\voffset by -1.0cm

\begin{document}

\title{CTI History of the EPIC pn Camera}

\author{K.\,Dennerl \and U.G.\,Briel \and M.J.\,Freyberg \and
        F.\,Haberl \and N.\,Meidinger \and V.E.\,Zavlin}

\institute{Max--Planck--Institut f\"ur extraterrestrische Physik,
           Giessenbachstra{\ss}e~1, 85748 Garching, Germany}

\maketitle 

\begin{abstract}

The pn camera of EPIC is inherently robust against radiation damage effects.
Nevertheless, the knowledge of charge transfer inefficiency (CTI) of the
pn camera is crucial for obtaining the correct energy scale. We describe
detailed in--orbit monitoring of this effect, utilizing the internal
calibration source. We find that during the first two years in orbit the CTI
increased by $\sim4\%$ for Al--K$_{\alpha}$ and by $\sim7\%$ for
Mn--K$_{\alpha}$. The increases in the CTI are well within expectations, with
no measurable effect on the energy resolution.


\end{abstract}

\begin{figure}[!ht]
\begin{center}
\includegraphics[clip,width=7.5cm]{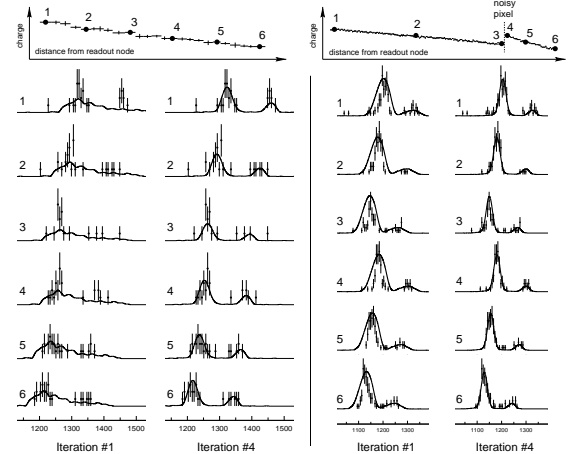}
\end{center}
\caption{Illustration of the charge loss determination by the template cross
correlation method, for the case of low (left) and high (right) statistical
quality. In both cases the two columns show the results of template fits to
spectra from selected macro pixels for the first (left) and last (right)
iteration. The macro pixels are identified by filled circles in the resultant
charge loss curves at top, where the apparent energy of the emission line is
plotted against the distance from the readout node. In the case at left only
631 events were recorded within the whole CCD column, leaving only 20\,--\,29
events for each of the 25 macro pixels. At right, 17\,864 events were
available in total, sufficient for appling this technique to each individual
pixel. Note how significantly the presence of a noisy pixel reduces the charge
loss for events which were shifted across this pixel during readout. In both
cases Mn--K$_{\alpha}$ and K$_{\beta}$ lines were analysed.
}
\label{kdennerl-WA2_fig:fig1}
\end{figure}

\section{Introduction}

In CCDs electrons of the signal charge generated by an absorbed
X--ray photon are lost during
transfer to the readout node by traps in the transfer channel. The number
of traps is expected to increase with time due to radiation damage.
As the knowledge of the charge loss is essential for extracting quantitative
spectral information, the performance of the EPIC pn camera is routinely
monitored with an internal calibration source. This source consists
of radioactive Fe$^{55}$ with an Al--target, and
irradiates the detector with Al--K$_{\alpha}$ and Mn--K$_{\alpha}$ emission
lines at 1.5 and 5.9~keV. This is usually done for about one hour at the
beginning of each 48~hour revolution of XMM--Newton. The line positions
determined from these exposures are a sensitive indicator for any change in
the energy response.

\section{Ground calibration}
\label{kdennerl-WA2:ground}

\begin{figure}[ht]
\begin{center}
\hbox{
\raisebox{1.68cm}{
\includegraphics[clip,bbllx=102pt,bblly=428pt,bburx=140pt,bbury=700pt,width=0.6cm]
                {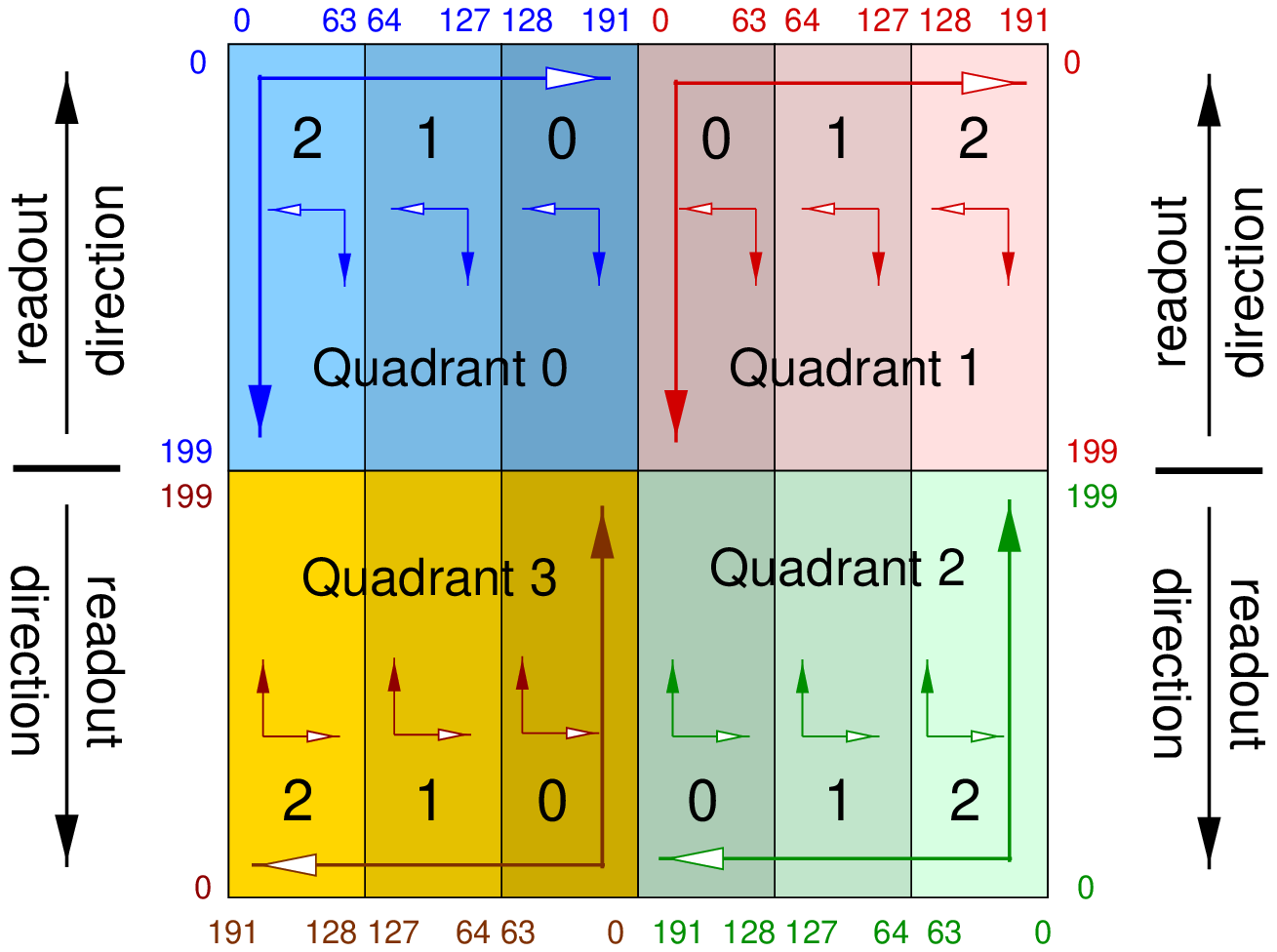}}
\includegraphics[clip,bbllx=45pt,bblly=20pt,bburx=580pt,bbury=640pt,width=7.8cm]
                {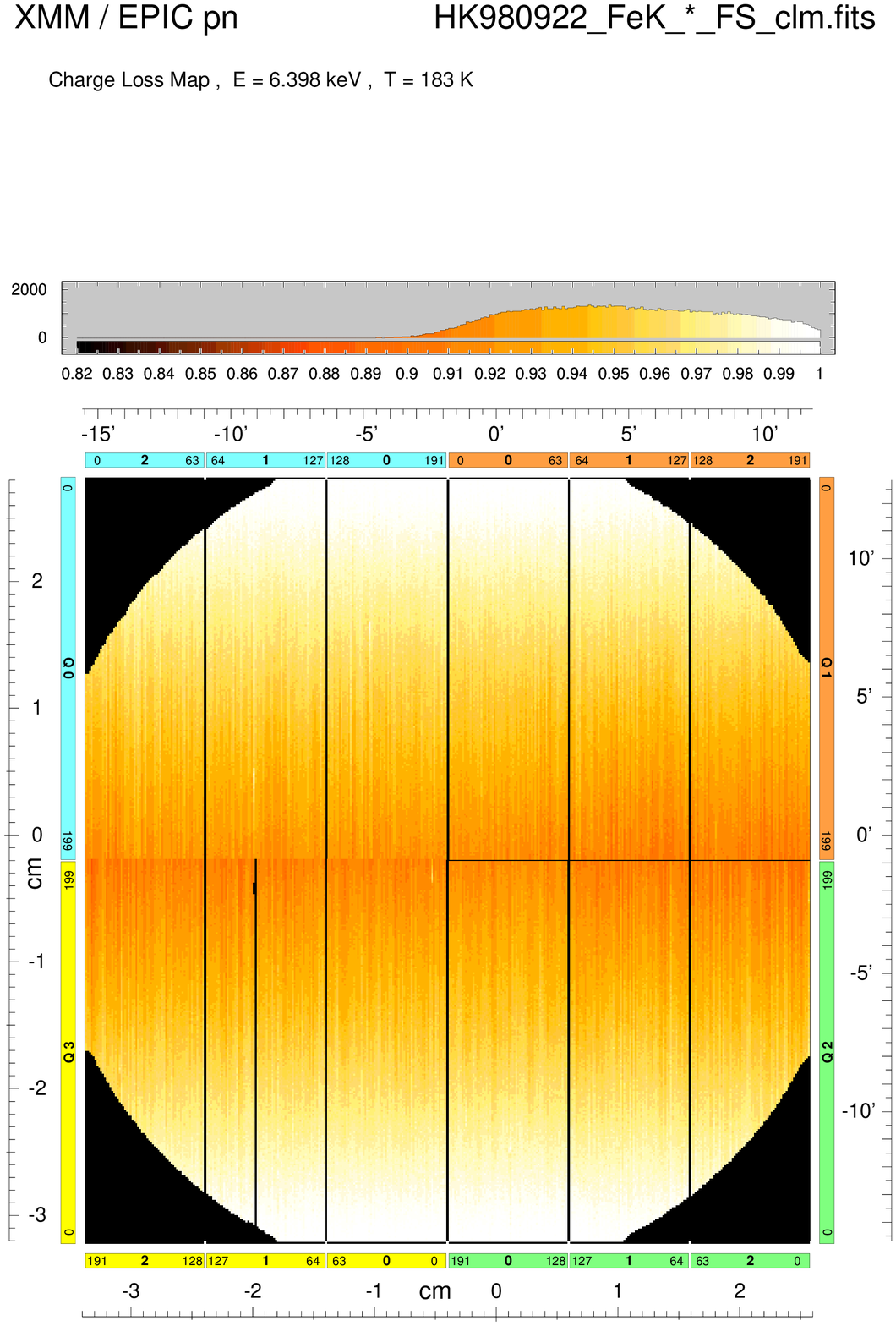}
}
\end{center}
\caption{Residual charge map at Fe--K$_{\alpha}$, derived from a superposition
of several long flatfield exposures during ground calibration at Orsay, where
more than 80 million events were recorded. The colour of each pixel indicates
the fraction of the charge which arives at the readout node, in the colour
coding which is displayed at top together with a frequency histogram.
Charge losses of up to 10\% occur, and by this amount the Fe--K$_{\alpha}$ line
energy would shift across the detector, if no CTI correction were applied.}
\label{kdennerl-WA2_fig:fig3}
\end{figure}

\begin{figure}[ht]
\begin{center}
\includegraphics[clip,bbllx=50pt,bblly=120pt,bburx=552pt,bbury=739pt,width=7.2cm]
                {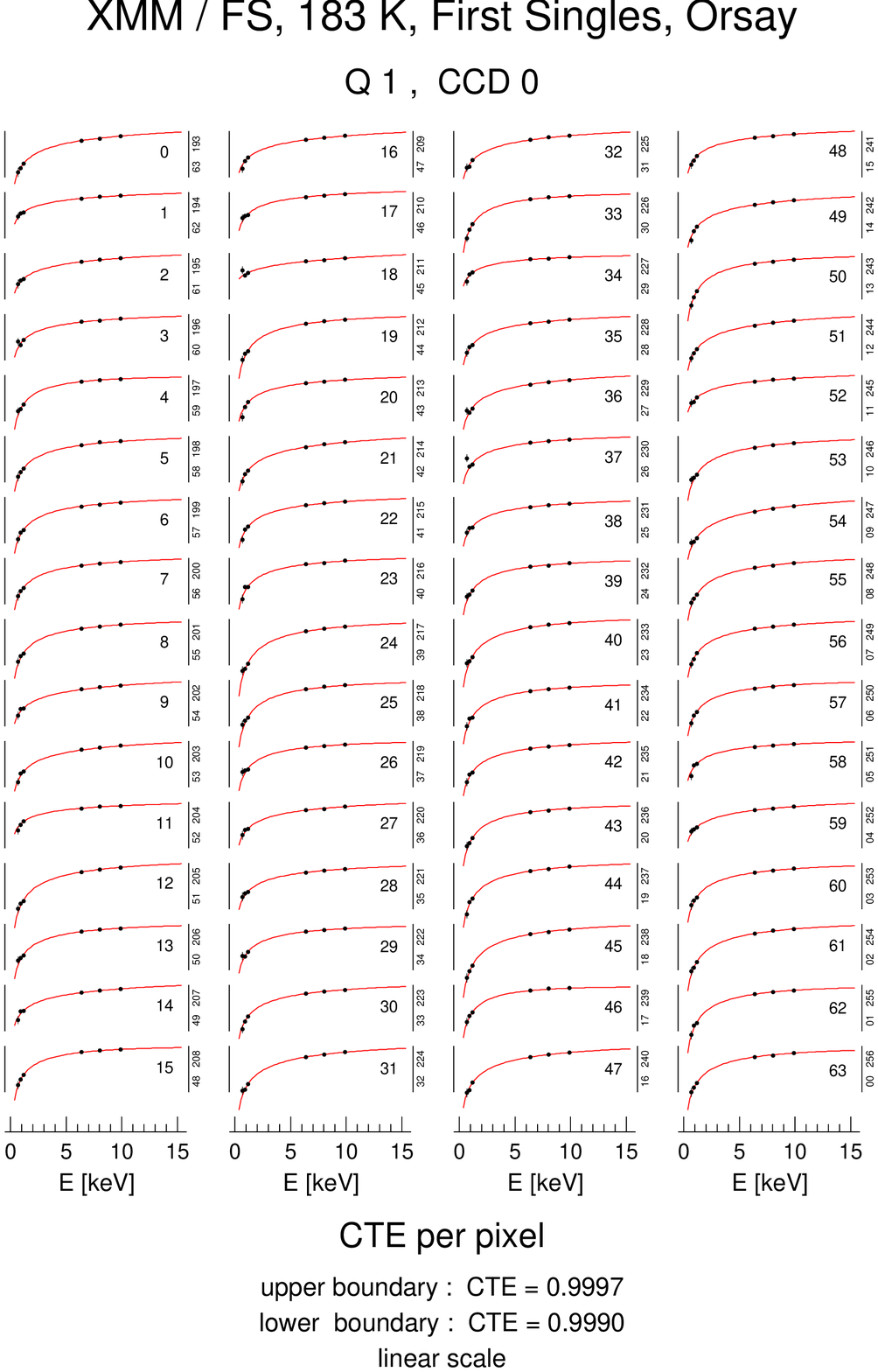}
\end{center}
\caption{Energy dependence of the CTE ($=1.0-\mbox{CTI}$)
for all columns of CCD\,4, as measured
before launch. Dots indicate the measurements, while the solid lines were
obtained by fitting the function $\mbox{CTE(E)}=a_0 + a_1\log E +
a_2\left(\log E\right)^2$ to the observed values. Vertical lines along the CTE
curves cover the range from 0.9990 to 0.9997 in a linear scale.}
\label{kdennerl-WA2_fig:fig4}
\end{figure}

The importance of the CTI for the energy calibration of the EPIC pn camera was
realized long before launch, and a lot of experience was gained during
extensive laboratory measurements with different detectors at different
temperatures and energies. We developed specific software tools for the CTI
analysis of such calibration data, which allow us to determine the charge loss
across the 768 readout channels with high spatial resolution.
Fig.\,\ref{kdennerl-WA2_fig:fig1} illustrates the method. Results from
laboratory measurements for the detector now onboard XMM--Newton are shown in
Figs.\,\ref{kdennerl-WA2_fig:fig3} and \ref{kdennerl-WA2_fig:fig4}. More about
the energy calibration before launch can be found in
\cite*{kdennerl-WA2:den99}. Here we report on recent results obtained from
monitoring the CTI in orbit.

\begin{figure}[ht]
\begin{center}
\includegraphics[clip,bbllx=35pt,bblly=25pt,bburx=565pt,bbury=765pt,width=7.5cm]
                {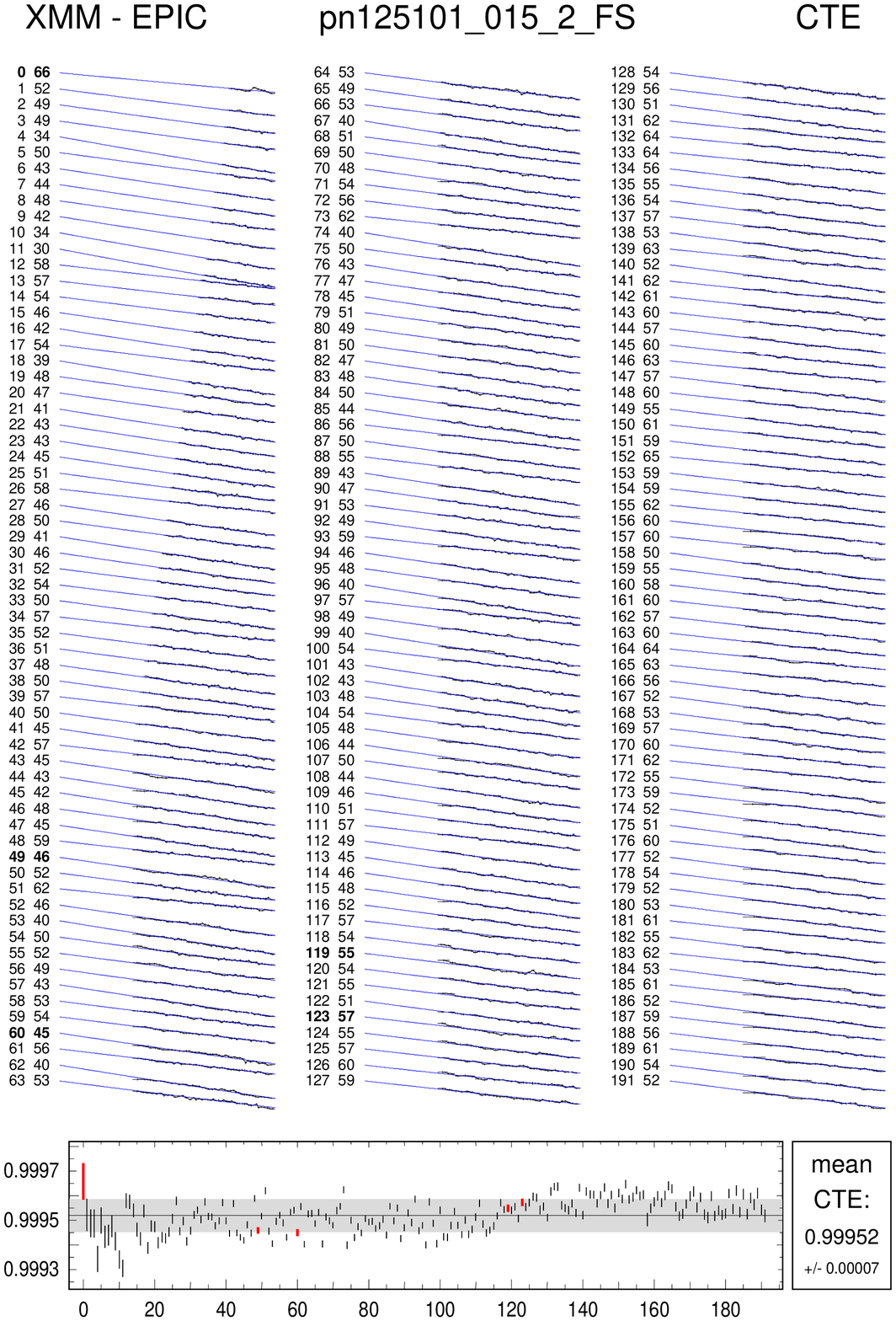}
\end{center}
\caption{Charge loss and CTE for individual columns of quadrant 2 at
Mn--K$_{\alpha}$, obtained from a 18~hour exposure with the internal
calibration source during revolution 125. The high statistical quality
made it possible to determine the charge losses for individual pixels.
The curves show the apparent line energy as a function of the distance
from the readout node. When shifted along the whole column, the charge
losses accumulate to $\sim10\%$. The first number at the curves identifies
the column, while the second number gives the CTE in the form
$\left(\mbox{CTE}-0.999\right)\cdot10^5$, i.e.\ the last two digits xx
of 0.999xx. The CTE was obtained by fitting an exponential function to the
charge losses. At bottom the CTE values of all columns are summarized.
Significant differences between the individual columns are obvious.
The shaded region shows the standard deviation of the CTE values, which
is printed to the right together with the mean value.
}
\label{kdennerl-WA2_fig:fig2}
\end{figure}

\section{CTI monitoring in orbit}
\label{kdennerl-WA2:orbit}

The number of photons obtained during the
$\sim1\mbox{ hour}$ calibration
measurements in orbit is not sufficient for the determination of the charge
loss for individual columns.
However, during six revolutions (\#\,23, 80, 125, 172,
242, and 332) long measurements were made with exposures of 13 up to 21~hours
duration, where the statistical quality is comparable to laboratory
measurements. Fig.\,\ref{kdennerl-WA2_fig:fig2} shows the charge losses
determined from rev.\,125, and Fig.\,7 in Briel et al.\ (2002, these
proceedings) summarizes the CTI history for CCDs 1, 2, and 3 (quadrant 0),
derived from the six measurements.

In order to extend the CTI analysis to the shorter exposures, we modified our
technique in the following way: we combined events, with the same
distance from the readout node, from several readout columns, after having
corrected their raw amplitudes for the gain of the particular column. As the
irradiation is not homogeneous across the detector
(Figs.\,\ref{kdennerl-WA2_fig:fig8} and \ref{kdennerl-WA2_fig:fig9}), we
selected only events from the better exposed areas, which are less contaminated
by out--of--time events. The spectra of such `macro columns' were then
analysed in the same way as those of individual columns
(cf.\,Fig.\,\ref{kdennerl-WA2_fig:fig1}). Fig.\,\ref{kdennerl-WA2_fig:fig5}
illustrates the accuracy which can be reached with this method.

\section{Results}
\label{kdennerl-WA2:results}

\begin{figure}[ht]
\begin{center}
\hbox to \hsize{\hfil
\includegraphics[clip,width=3.7cm]{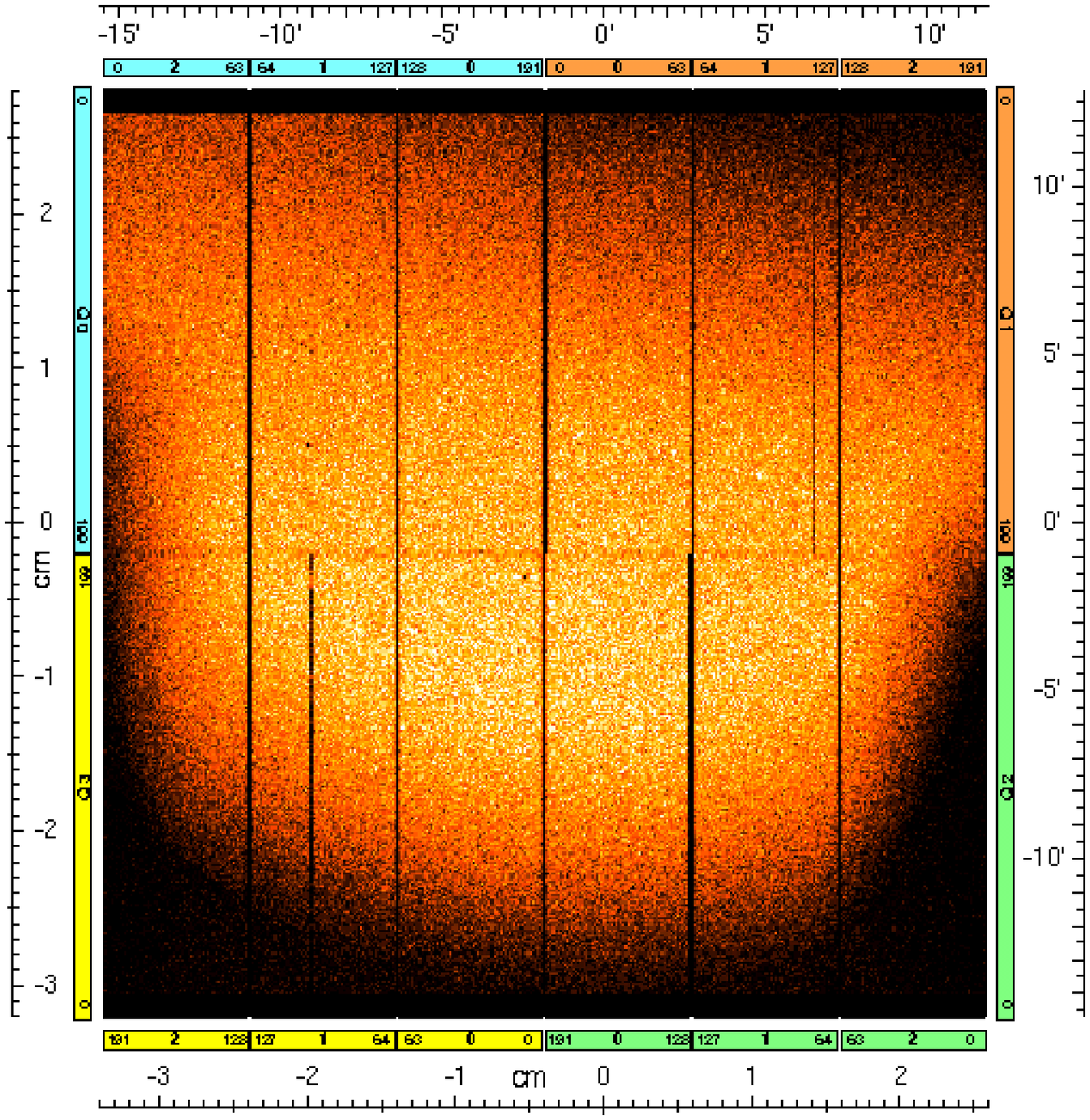}
\hfil
\includegraphics[clip,width=3.7cm]{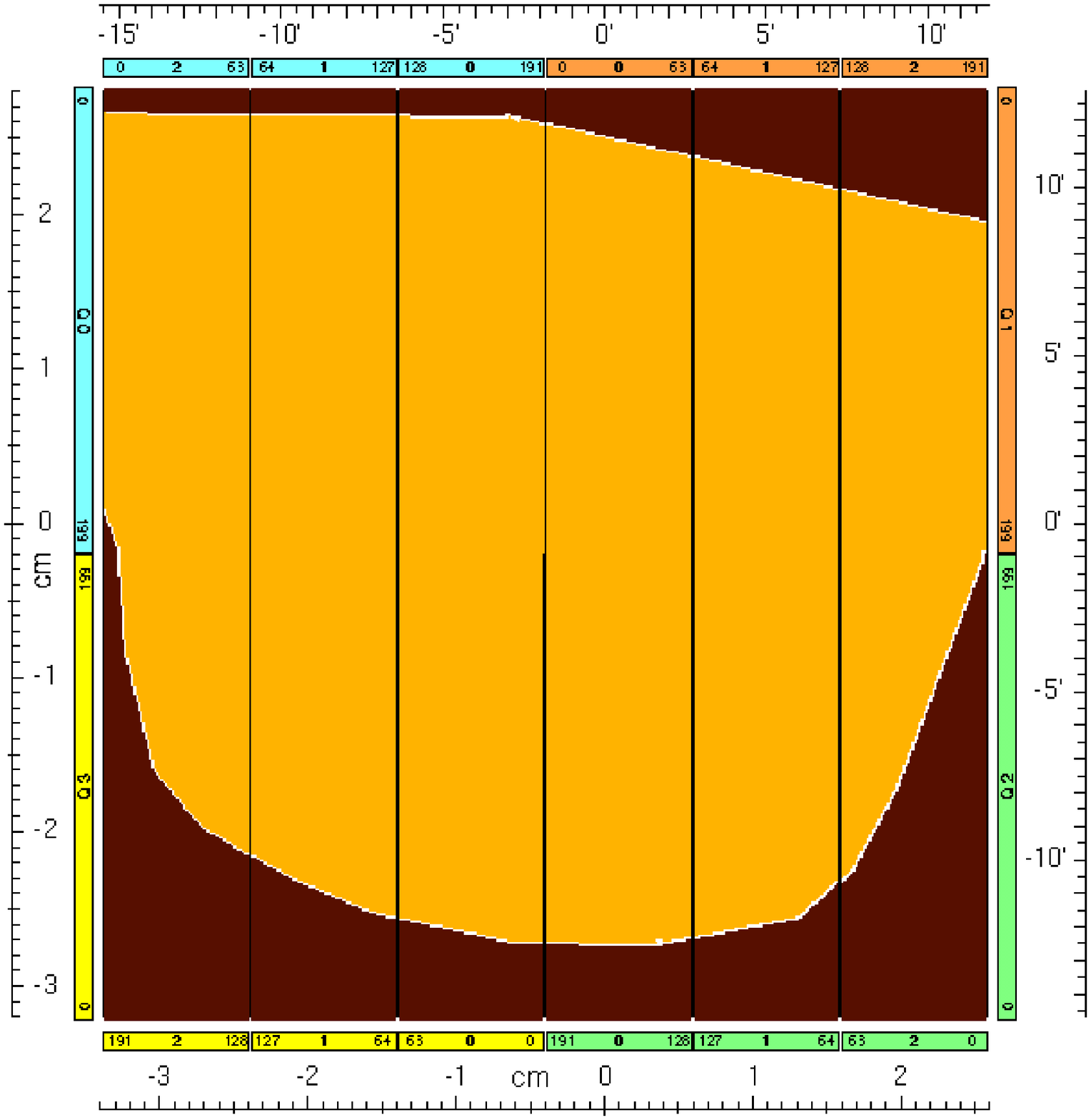}
\hfil}
\end{center}
\vspace*{-4mm}
\caption{Irradiation at Al--K$_{\alpha}$ and corresponding mask.}
\label{kdennerl-WA2_fig:fig8}
\end{figure}

\begin{figure}[ht]
\begin{center}
\hbox to \hsize{\hfil
\includegraphics[clip,width=3.7cm]{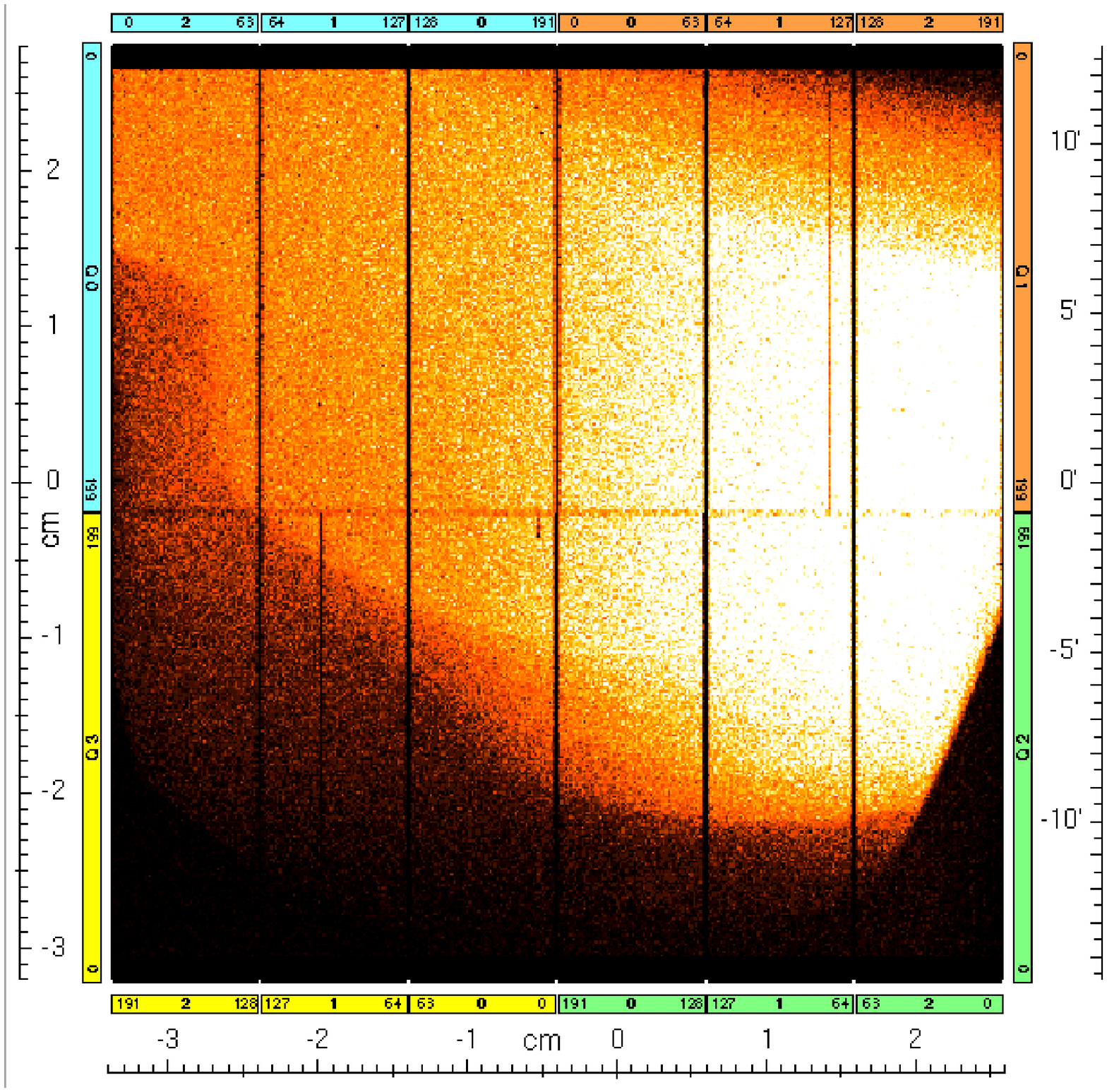}
\hfil
\includegraphics[clip,width=3.7cm]{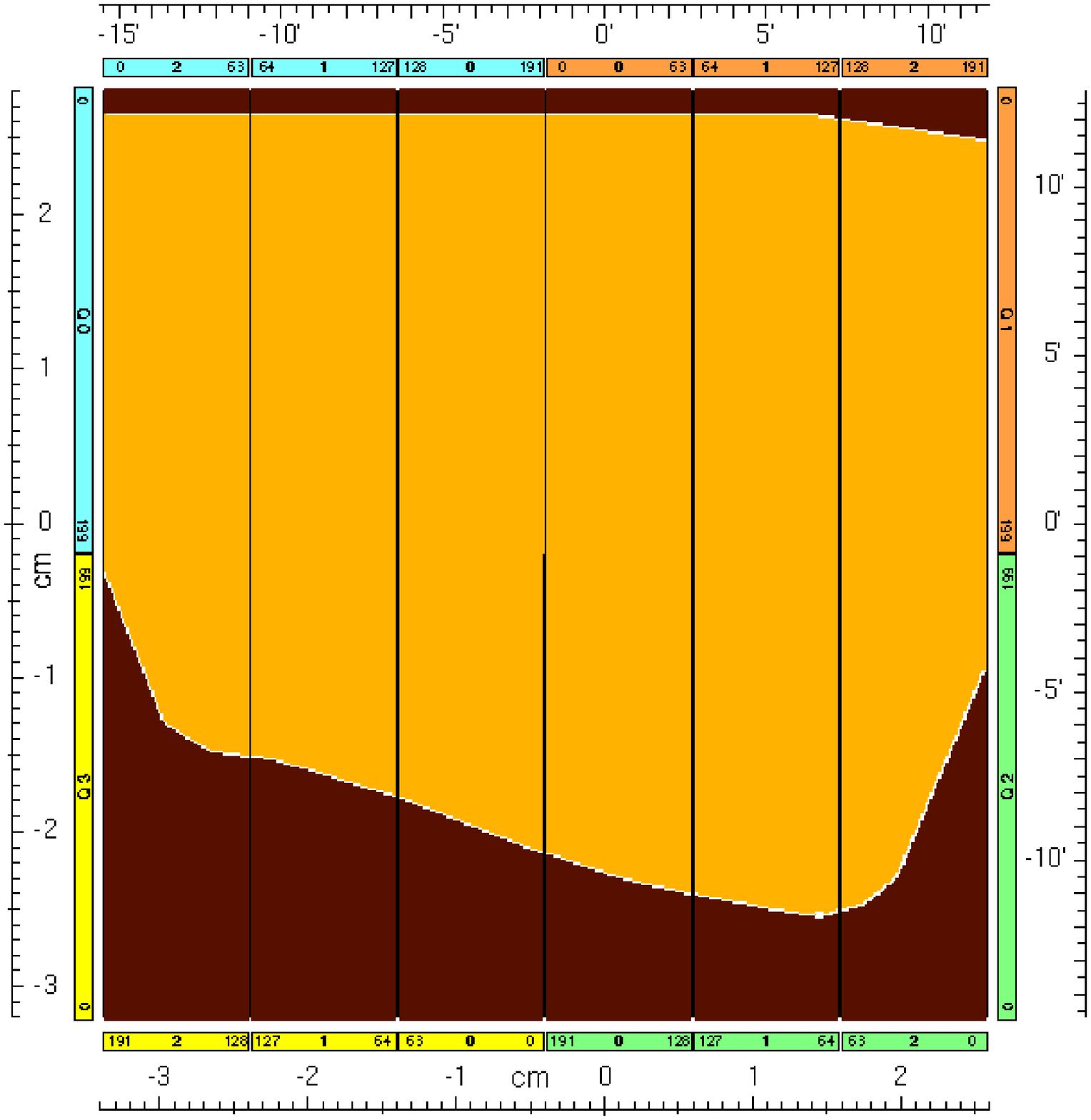}
\hfil}
\end{center}
\vspace*{-4mm}
\caption{Irradiation at Mn--K$_{\alpha}$ and corresponding mask.}
\label{kdennerl-WA2_fig:fig9}
\end{figure}

\begin{figure}[!ht]
\begin{center}
\includegraphics[clip,bbllx=45pt,bblly=264pt,bburx=565pt,bbury=765pt,width=7.5cm]
                {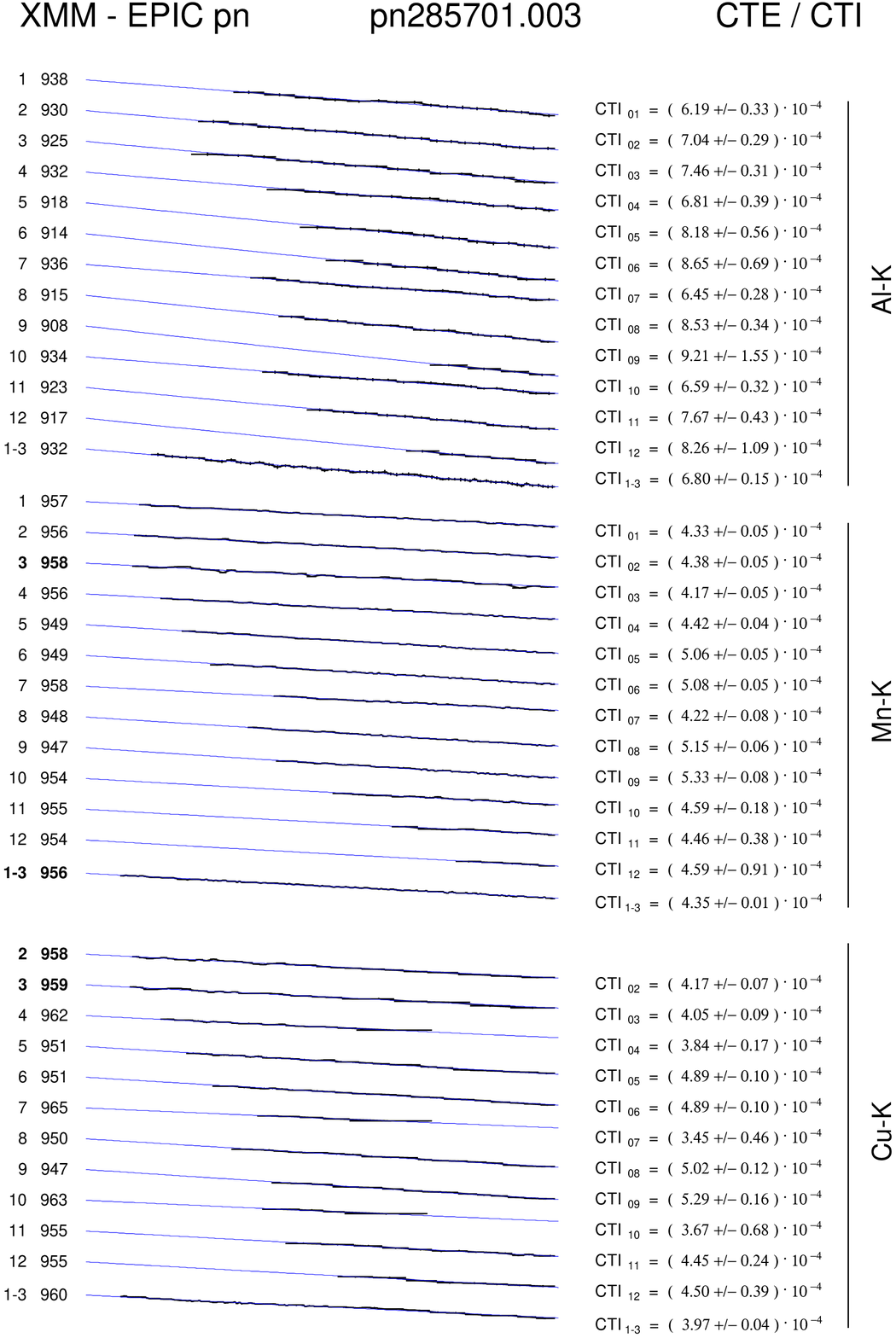}
\end{center}
\caption{Charge loss and CTI for individual CCDs, and for quadrant~0. This
plot is similar to Fig.\,\ref{kdennerl-WA2_fig:fig2}. Here, however, photons
from several columns were combined (after having been corrected for individual
gain variations), to increase the statistical quality and to extend the CTI
analysis to shorter exposures. The data were taken from a 2~hour exposure with
the internal calibration source during revolution 285. The curves refer to
CCDs 1\,--\,12, and to quadrant~0, for Al--K$_{\alpha}$ (top) and
Mn--K$_{\alpha}$ (bottom). Gaps are caused by poorly exposed regions
(cf.\,Fig.\,\ref{kdennerl-WA2_fig:fig8}, \ref{kdennerl-WA2_fig:fig9}). The CTI
values, determined from exponential fits, are listed to the right of each
curve.}
\label{kdennerl-WA2_fig:fig5}
\end{figure}

\begin{figure}[ht]
\begin{center}
\includegraphics[clip,bbllx=40pt,bblly=100pt,bburx=540pt,bbury=350pt,width=8.8cm]
                {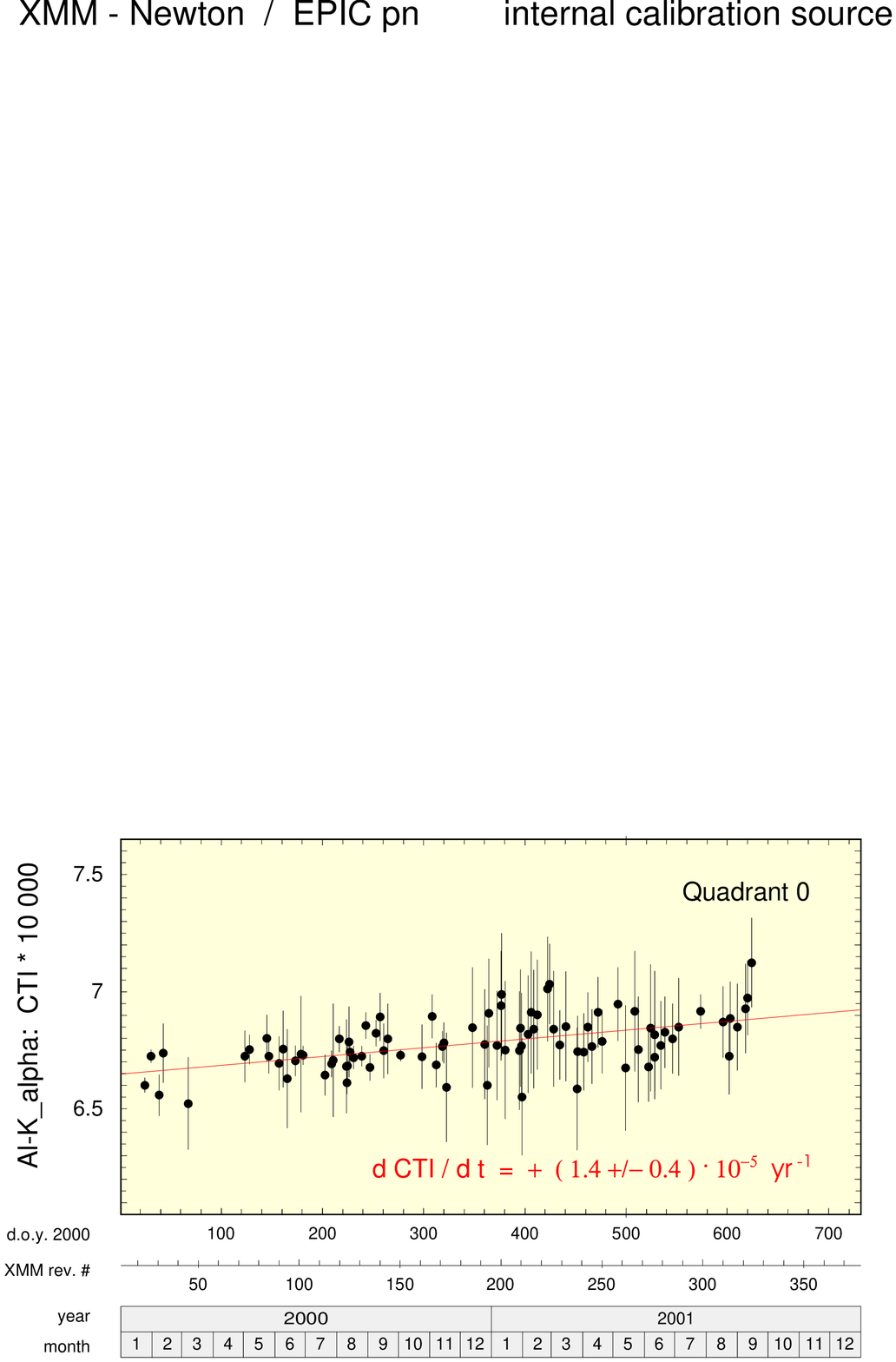}
\includegraphics[clip,bbllx=40pt,bblly=25pt,bburx=540pt,bbury=340pt,width=8.8cm]
                {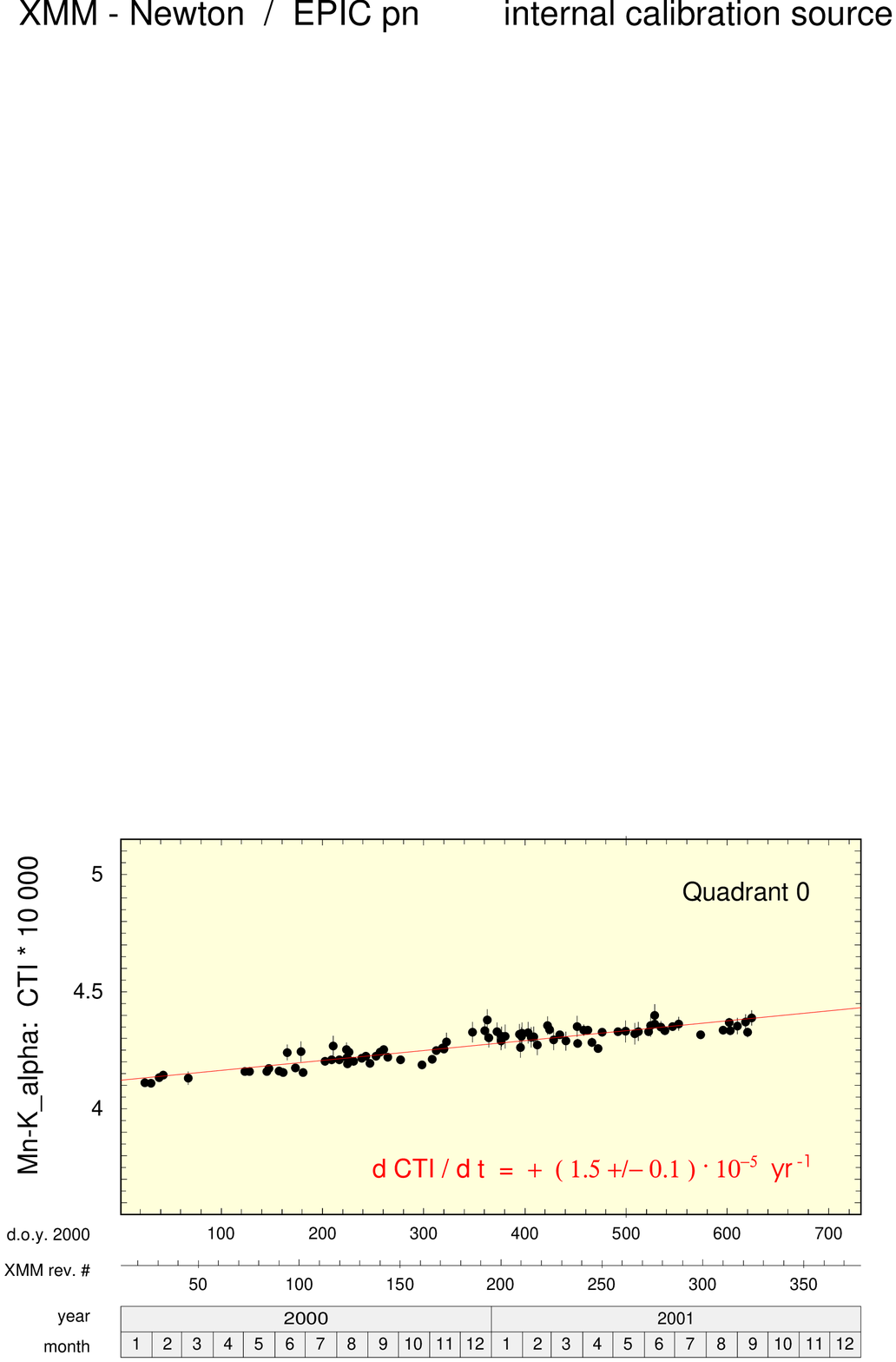}
\end{center}
\caption{Results of the CTI determination from internal calibration
measurements in orbit, for quadrant~0 at Al--K$_{\alpha}$ (top) and
Mn--K$_{\alpha}$ (bottom). Both energies show a consistent slope, which
is significantly different from zero in both cases.
}
\label{kdennerl-WA2_fig:fig10}
\end{figure}

Fig.\,\ref{kdennerl-WA2_fig:fig10} summarizes the CTI
results for Al--K$_{\alpha}$ and Mn--K$_{\alpha}$ obtained for quadrant~0,
which receives
relatively homogeneous irradiation (Figs.\,\ref{kdennerl-WA2_fig:fig8},
\ref{kdennerl-WA2_fig:fig9}). We find the CTI at Al--K$_{\alpha}$ and
Mn--K$_{\alpha}$ to
increase slowly with time, with a rate of
$\sim1.5\cdot10^{-5}\mbox{ yr}^{-1}$ for both energies.

\begin{figure}[!ht]
\begin{center}
\includegraphics[clip,bbllx=40pt,bblly=460pt,bburx=540pt,bbury=775pt,width=8.8cm]
                {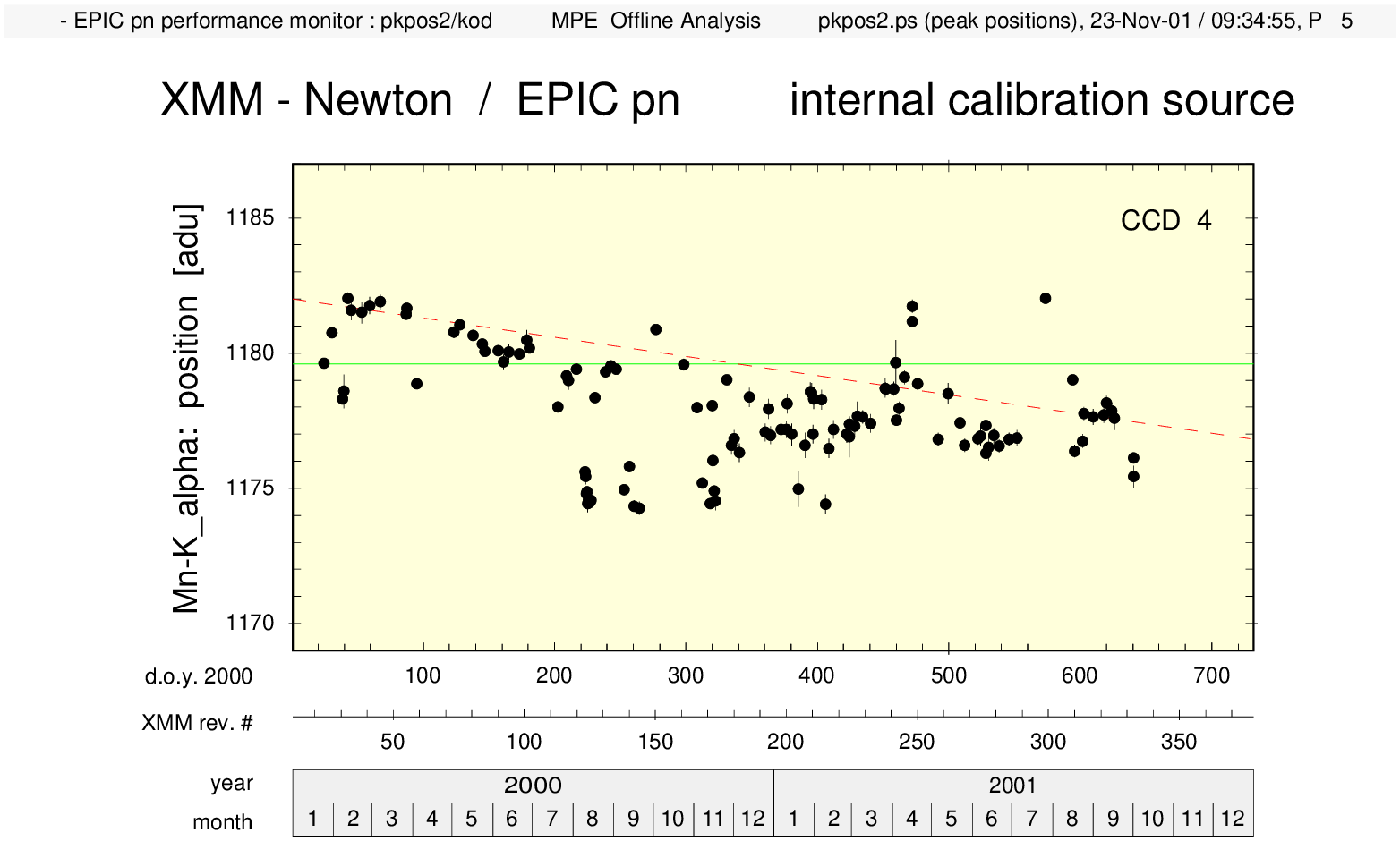}
\end{center}
\caption{Mn--K$_{\alpha}$ line position, not corrected for CTI history. In the
first months after the launch of XMM--Newton, short--term drops of the peak
position on time scale of days showed up, caused by temperature variations in
the electronic boxes, while brief, sudden rises were found to be related to
episodes of very high background. In addition to these short--term changes
there are indications for a long--term decrease of the line position (dashed
red line), which is caused by a CTI increase. The green line marks the nominal
value of Mn--K$_{\alpha}$ for an amplification of 5~{\rm eV}/adu.}
\vspace*{0.4cm}
\label{kdennerl-WA2_fig:fig7}
\end{figure}

This is in contrast to the considerable scatter seen in the Mn--K$_{\alpha}$
line positions, which result when we apply the same CTI correction to the
calibration data (Fig.\,\ref{kdennerl-WA2_fig:fig7}). We found that the
short--term drops of the line position were not caused by CTI changes, but
by changes in the amplification, which were the consequence of temperature
variations in the electronic boxes. The drops in the line position
show a clear correlation with the temperature and can thus be well
corrected. Short--term rises of the line position, on the other hand, were
found to be correlated with periods when the background radiation was so high
that no astrophysical observations could be performed. Despite all the scatter,
however, Fig.\,\ref{kdennerl-WA2_fig:fig7} shows evidence for a long--term
decrease of the line position, which is indeed caused by the increase of the
CTI.

How does the CTI increase seen during the first two years in space compare
with pre--launch predictions\,? \,For Mn--$K_{\alpha}$ we measure a long-term trend of
$$ {d\,{\mit CTI}\over dt} = +\left(1.5 \pm 0.1\right) \cdot
                             10^{-5}\mbox{ yr}^{-1}. $$
Before launch, laboratory measurements, taken at $-90^{\circ}\mbox{ C}$,
the operating temperature of the EPIC pn camera in orbit,
showed the following response of the CTI to a
10~MeV proton equivalent flux $F_{\rm rad}$ for Mn--K$_{\alpha}$
(\cite{kdennerl-WA2:mei98}):
$$ {d\,{\mit CTI}\over dt} = 4\cdot10^{-13}\mbox{ cm}^2 \cdot F_{\rm rad} $$
For $F_{\rm rad}$, an average value of
$5\cdot10^7\mbox{ cm}^{-2}\mbox{ yr}^{-1}$ was expected, yielding
$$ {d\,{\mit CTI}\over dt} =
   +\,2\cdot10^{-5}\mbox{ yr}^{-1} $$
at Mn--K$_{\alpha}$. Thus, the measured value is even somewhat lower than
this estimate.

\begin{figure}[ht]
\vspace*{3.2mm}
\begin{center}
\includegraphics[clip,bbllx=40pt,bblly=460pt,bburx=540pt,bbury=765pt,width=8.8cm]
                {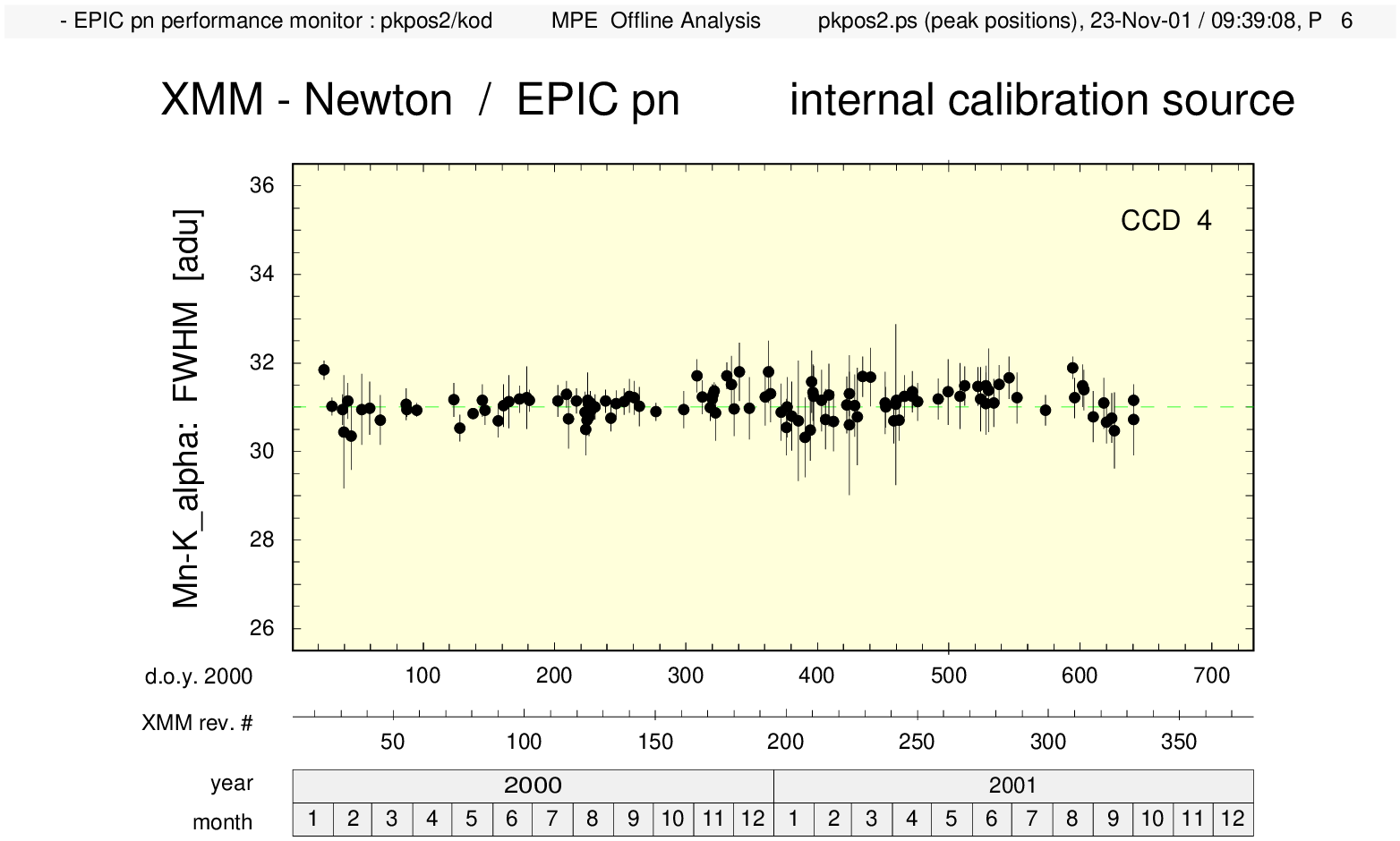}
\end{center}
\caption{Energy resolution for Mn--K$_{\alpha}$. The FWHM of 31~adu corresponds
to 155~{\rm eV}. During the first two years in orbit, no change in the energy
resolution was observed.}
\vspace*{2mm}
\label{kdennerl-WA2_fig:fig6}
\end{figure}

The additional noise created by the small relative increase of the CTI is
almost negligible and should not have measurable consequences for the energy
resolution. In fact, no significant increase of the width of the
Mn--K$_{\alpha}$ line was observed over the first two years
(Fig.\,\ref{kdennerl-WA2_fig:fig6}).

\section{Summary and conclusions}
\label{kdennerl-WA2:summary}

From routine measurements with the internal calibration source, the CTI
of the EPIC pn camera was found to increase by
\vspace*{-5mm}
$$+\left(1.4\pm0.4\right)\cdot10^{-5}\mbox{ yr}^{-1}
   \quad\mbox{at Al--K$_{\alpha}$,}$$
\vspace*{-4mm}\noindent
and by
\vspace*{-3.25mm}
$$+\left(1.5\pm0.1\right)\cdot10^{-5}\mbox{ yr}^{-1}
   \quad\mbox{at Mn--K$_{\alpha}$.}$$

\vspace*{-0.7mm}\noindent
This corresponds to a relative increase of the CTI by $\sim4\%$ for
Al--K$_{\alpha}$ and by $\sim7\%$ for Mn--K$_{\alpha}$ during the first two
years in orbit, with no measurable effect on the energy resolution. If this
trend continued, then it would take more than 25 years until the CTI at
Mn--K$_{\alpha}$ would have doubled. At Al--K$_{\alpha}$, the CTI would then
have increased by about half of its present value.

\vspace*{5mm}
\begin{acknowledgements}

The XMM--Newton project is an ESA Science Mission with instruments and
contributions directly funded by ESA Member States and the USA (NASA).
The XMM--Newton project is supported by the Bundesministerium f\"ur
Bildung und For\-schung / Deutsches Zentrum f\"ur Luft- und Raumfahrt
(BMBF /\,DLR), the Max--Planck--Gesellschaft and the Heidenhain--Stiftung.

\end{acknowledgements}

\end{document}